\documentclass[useAMS,usenatbib]{mn2e}

\usepackage{graphics}

\title[Radio jets of ULX Holmberg~II~X-1]{Unveiling recurrent jets of the ULX Holmberg~II~X-1: evidence for a massive stellar-mass black hole?}
\author[D. Cseh et al.]{D. Cseh$^{1}$\thanks{E-mail:
d.cseh@astro.ru.nl}, P. Kaaret$^{2}$, S. Corbel$^{3}$, F. Gris\'{e}$^{4,5}$, C. Lang$^{2}$, E. K\"{o}rding$^{1}$, H. Falcke$^{1}$, 
\newauthor P. G. Jonker$^{6,1,7}$, J. C. A. Miller-Jones$^{8}$, S. Farrell$^{9}$, Y. J. Yang$^{10}$, Z. Paragi$^{11}$, S. Frey$^{12}$
\\
$^{1}$Department of Astrophysics/IMAPP, Radboud University Nijmegen, P.O. Box 9010, 6500 GL Nijmegen, The Netherlands\\
$^{2}$Department of Physics and Astronomy, University of Iowa, Iowa City, 52240, USA\\
$^{3}$Laboratoire AIM (CEA/IRFU-CNRS/INSU-Universit\'{e} Paris Diderot), CEA/DSM/IRFU/SAp, F-91191 Gif-sur-Yvette, France\\
$^{4}$Instituto de Astrof\'isica de Canarias, E-38200 La Laguna, Tenerife, Spain\\
$^{5}$Departamento de Astrof\'isica, Universidad de La Laguna, Avda. Astrof\'isico Francisco Sanchez s/n, E-38271 La Laguna, Tenerife, Spain\\
$^{6}$SRON, Netherlands Institute for Space Research, Sorbonnelaan 2, 3584 CA Utrecht, The Netherlands\\
$^{7}$Harvard--Smithonian Center for Astrophysics, 60 Garden Street, Cambridge, MA~02138, USA\\
$^{8}$International Centre for Radio Astronomy Research - Curtin University, GPO Box U1987, Perth, WA 6845, Australia\\
$^{9}$Sydney Institute for Astronomy, School of Physics, The University of Sydney, NSW 2006, Australia\\
$^{10}$Astronomical Institute, Anton Pannekoek, University of Amsterdam, Postbus 94249, 1090 GE Amsterdam, The Netherlands\\
$^{11}$Joint Institute for VLBI in Europe, Postbus 2, 7990 AA Dwingeloo, The Netherlands\\
$^{12}$F\"{O}MI Satellite Geodetic Observatory, P.O. Box 585, H-1592 Budapest, Hungary\\
}

\begin{document}
\date{Draft}
\pagerange{\pageref{firstpage}--\pageref{lastpage}} \pubyear{2013}
\maketitle
\label{firstpage}

\begin{abstract}
We report on the discovery of an apparent triple radio structure hidden inside the radio bubble of the ultraluminous X-ray source Holmberg II X-1. The morphology is consistent with a collimated jet structure, which is observed to emit optically thin synchrotron radiation. The central component has a steep radio spectrum and is brighter than the outer components indicating a renewed radio activity. We estimate a minimum time-averaged jet power of $\sim$$2 \times 10^{39}$~erg~s$^{-1}$ that is associated with a time-averaged isotropic X-ray luminosity of at least $4\times10^{39}$~erg~s$^{-1}$. Our results suggest that Holmberg~II~X-1 is powered by a black hole of $M_{\rm{BH}}\geq25$~M$_{\odot}$, that is inferred to be accreting at a high Eddington rate with intermittent radio activity.
\end{abstract}


\section{Introduction}
 
A population of extragalactic black holes (BHs), called ultraluminous X-ray sources (ULXs), possess extreme power output ($L_{\rm{x}}>3 \times 10^{39}$ erg s$^{-1}$) compared to stellar-mass BHs ($M_{\rm{BH}}=3-20$~M$_{\odot}$), primarily in the form of radiation that is released via an accretion disc formed from infalling gas from a companion star. Holmberg~II~X-1 (hereafter Ho~II~X-1) lies at a distance of 3.39 Mpc \citep{Karachentsev:2002fk} in a dwarf irregular galaxy. Such galaxies are often thought to mimic the early cosmological conditions due to their low metallicities. In these environments, massive stellar-mass BHs ($M_{\rm{BH}}=20-100$~M$_{\odot}$) are predicted to form \citep[e.g.][]{Fryer:1999fk,Zampieri:2009if,Belczynski:2010kx,Mapelli:2013uq}. However, there has been little or no observational evidence for this class of BHs, beyond IC~10~X-1 \citep[e.g.][]{Bauer:2004sh,Prestwich:2007kx,Silverman:2008uq} and NGC~300~X-1 \citep[e.g.][]{Crowther:2010fk}. Studying potential local analogues can have implications for understanding BH seed formation and growth, see the review by \citet{Volonteri:2012uq}.     

Inferred from X-ray spectroscopic studies \citep{Goad:2006fk} and from numerical simulations \citep{Mapelli:2007uq}, the maximum mass of Ho~II~X-1 was estimated to be 100~M$_{\odot}$ and $10^{3}-10^4$~M$_{\odot}$, respectively. Also, from the predicted scale invariance of the accretion variability timescale and of the jets \citep{McHardy:2006zk,Merloni:2003ri,Falcke:2004wm}, it might be expected that a massive stellar-mass BH would show radio activity on a longer timescale compared to microquasars. Furthermore, such powerful BHs may fundamentally regulate their environment in the form of both jets and X-ray ionisation \citep[e.g.][]{Pakull:2002fk,Gallo:2005wa,Pakull:2010kq,Cseh:2012fk,Mezcua:2013fk}. Due to the potential presence of a higher mass BH and longer active cycles, these effects could be more pronounced in ULXs than in XRBs hosting stellar-mass BHs. If this is indeed the case, ULXs may be relevant at the epoch of reionization in the early Universe \citep{Mirabel:2011kx,Justham:2012vn}.

In this paper we present radio observations of Ho II X-1 that have a higher angular resolution and more sensitive than previous studies. Our observations offer a unique opportunity to estimate the time-averaged jet power in a ULX that is dominated by photoionisation and not shock-ionization \citep{Lehmann:2005bu,Egorov:2013ys}.This allows us to test the coupling between the jet power and the previously estimated time-averaged bolometric luminosity \citep{Kaaret:2004ta,Berghea:2010vn} and to infer a lower limit on the BH mass.

\section{Observation, Analysis, and Results}

We obtained radio observations of Ho~II~X-1 using the Karl G. Jansky Very Large Array (JVLA) on 16~Nov and 17~Nov~2012 under project code 12B-037 (PI Cseh). The observations were scheduled between 12:50-16:20~UT and 12:00-15:30~UT, respectively. The total on-source time was 5.25~hour in A~configuration, C~band, with a correlator integration time of 1~s. The total bandwidth was 2~GHz, covering 4.5--6.5~GHz, that consists of two base bands with 8-bit sampling, each with 8 sub-bands (or spectral windows) with 4 polarisation products, 64 channels per sub-band with 2~MHz channel width. 

The data were calibrated and imaged using CASA version 4.1.0 following standard procedures. We used 3C286 for absolute flux density and bandpass calibration, and J0841+7053 was used for antenna gain and phase calibration. At each epoch, we observed the primary calibrator for 10 min and the phase calibrator for 2 min out of every 10 min. The last four spectral windows were heavily affected by radio frequency interference and were flagged. We note that no self-calibration was applied.

\begin{figure}
\resizebox{3.5in}{!}{
\includegraphics{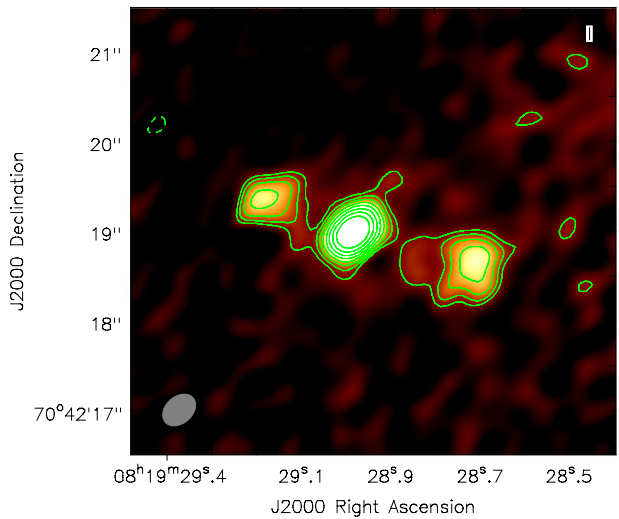}}
\caption{The C-band JVLA A-array image of Ho~II~X-1. The radio image was made using Briggs (robust=0.5) weighting. The lowest contours represent radio emission at $\pm(\sqrt{2})^{3}$ times the rms noise level of 2.5~$\mu$Jy beam$^{-1}$. Contours are increased by $(\sqrt{2})^{n}$$\sigma$, where $n\in[4;10]$, and the peak brightness is 151~$\mu$Jy~beam$^{-1}$.  The Gaussian restoring beam size is $0.41" \times 0.30"$ at a major axis position angle of --48$^\circ$. (The background colour image, with smoothed pixels, represents radio emission on an arbitrary scale of $0-30$~$\mu$Jy to match the contours.)}
\label{highres}
\end{figure}

\begin{figure}
\resizebox{3.5in}{!}{
\includegraphics{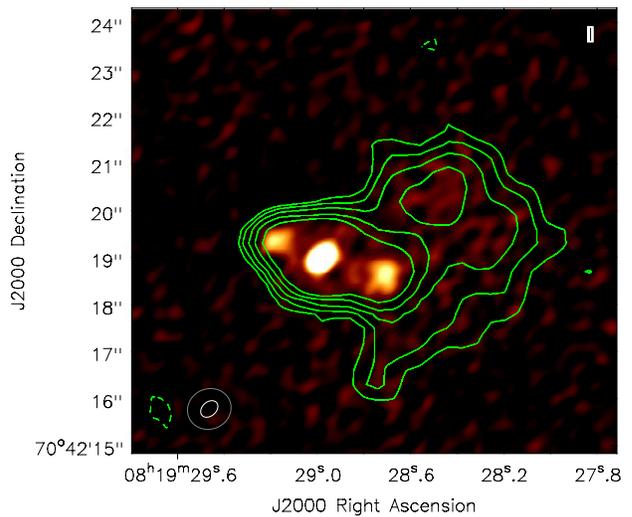}}
\caption{The low-resolution, naturally weighted, C-band A-array image. The figure shows the radio bubble associated with Ho~II~X-1 together with the fine-scale structure of Fig. \ref{highres} shown as background colour image. The lowest contours represent radio emission at $\pm(\sqrt{2})^{3}$ times the rms noise level of 3.2~$\mu$Jy beam$^{-1}$. Contours are increased by $(\sqrt{2})^{n}$$\sigma$, where $n\in[4;7]$, and the peak brightness is 192~$\mu$Jy~beam$^{-1}$. The radio image was made using natural weighting. The Gaussian restoring beam size is $0.94" \times 0.86"$ at a position angle of --68$^\circ$. The smaller beam (bottom left corner, in white) corresponds to the beam in Fig. \ref{highres}.}
\label{nebula}
\end{figure}

Images were made using the multi-frequency synthesis method, either using Briggs weighting with robust=0.5 (Fig. \ref{highres}) or using natural weighting (Fig. \ref{nebula}). First we imaged the data corresponding to the individual epochs in order to search for variability. As a next step, we concatenated the two data sets to achieve a better noise level. To bring out the already known, large-scale structure of the radio bubble \citep{Miller:2005xh,Cseh:2012fk}, we restricted the UV range to $<200$~k$\lambda$ to produce Fig. \ref{nebula}. By this restriction, the apparent maximum baseline length is one fourth of the original data set and the excluded fraction of visibilities is about 48\%. This allowed us to have a synthesised beam width 4 times larger -- i.e. comparable to the angular resolution of previous observations \citep{Cseh:2012fk} -- and to detect emission from larger spatial scales. 

We obtained integrated flux densities ($S_{\nu}$) and peak brightness intensities by fitting single-component Gaussians on the Briggs-weighted images using the AIPS task {\sc jmfit}. A summary of our results is available in the Online Material (OM). We find that the central component is point-like, and the north-west (NW) and south-east (SE) components are resolved. To estimate the spectral index of the central component, we imaged the first four ($\nu=4.75$~GHz) and the third four ($\nu=5.75$~GHz) spectral windows individually. Given that the central component is point-like, we use the peak intensity values to estimate a spectral index and we find $\alpha_{\rm{C}} =-0.8 \pm 0.2$, where $S_{\nu}\sim \nu^{\alpha}$; $\nu$ denotes the frequency. Using flux densities would result in a consistent spectral index of $\alpha_{\rm{C}} =-0.8 \pm 0.3$ with a larger uncertainty. The nominal spectral index of the SW and the NE components are $\alpha_{\rm{SW}} =-1.0 \pm 1.3$ and $\alpha_{\rm{NE}} =-1.0 \pm 1.4$

The fitted position of the central component corresponds to 8$^h$19$^m$28$^s$.9818 $\pm$ 0$^s$.0005, +70$^{\circ}$~42'~18".992 $\pm$ 0".002.  This is well within the positional error of the Chandra and HST counterpart \citep{Kaaret:2004ta}. The positions of the SW and the NE components are 8$^h$19$^m$28$^s$.718 $\pm$ 0$^s$.005, +70$^{\circ}$~42'~18".62 $\pm$ 0".02 and 8$^h$19$^m$29$^s$.165 $\pm$ 0$^s$.005, +70$^{\circ}$~42'~19".35 $\pm$ 0".02, respectively. The angular distance between the NE and SW components is 2.34 arcsec, corresponding to a projected spatial scale of 38.5 pc at a distance of 3.39 Mpc. The angular distance between the central and SW components is 1.36 arcsec whilst the distance between the central and NE components is 0.98 arcsec. We place an upper limit on the intrinsic size of the central component of $<3.9$ pc using the Gaussian restoring beam size of a uniformly weighted image of $0.28" \times 0.19"$.

Owing to the resolution and sensitivity of the JVLA, we resolved the inner structure of the associated radio bubble into a triple radio source (Fig. \ref{highres}). We detected a central component with a flux density of $S_{\nu}=174\pm4$~$\mu$Jy, and a SW and NE component with $S_{\rm{SW}}=85\pm9$~$\mu$Jy and $S_{\rm{NE}}=55\pm7$~$\mu$Jy at $>9 \sigma$ level. The SW and NE components appear to be extended with an average intrinsic diameter of $545\pm60$~mas (corresponding to 9.0 pc) and $405\pm60$~mas (corresponding to 6.7 pc). The central component is unresolved with a size of $<3.9$~pc. The projected distance between the outer components is 38.5 pc. The apparent morphology of the triple source shows a collimated jet structure. 

Comparing the two epochs of observations, we find that there is no evidence of significant variability of the central component that may argue against strong Doppler boosting (see also later). We find that the central component has a steep radio spectral index of $\alpha=-0.8\pm0.2$ between 4.75 and 5.75 GHz that is consistent with optically thin synchrotron emission. Furthermore, this component is brighter with respect to the outer SW and NE components by at least a factor of 2-3. The steep spectral index and the relative brightness of the component argues against a self-absorbed compact jet that is typically associated with an inefficient accretion state \citep{Fender:2009bh}. Instead, it is consistent with optically thin compact ejecta.

\section{Discussion}

\begin{figure}
\resizebox{3.35in}{!}{
\includegraphics{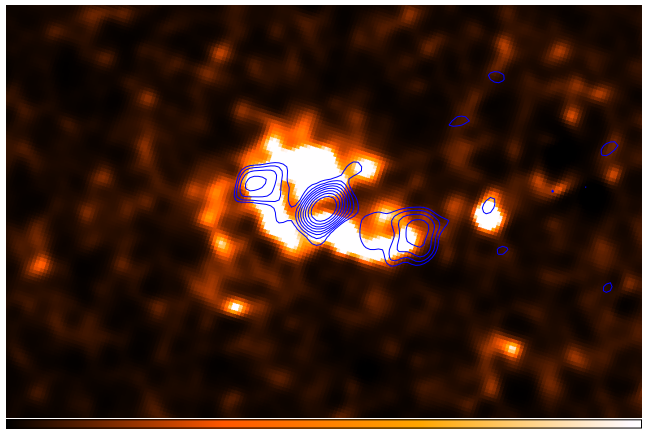}}
\caption{Radio contours from Fig. \ref{highres}. are overlaid on a HST image of \citet{Kaaret:2004ta} that is centred on the He {\sc ii} $\lambda4686$ line using the FR462N narrow-band filter.}
\label{overlay}
\end{figure}

\subsection{Kinematics}

The distance from the central component to the brighter\footnote{Assuming that the component does not appear brighter due to confusion with larger scale emission.} SW component is larger than that to the fainter NE component. A smaller arm-length to the east could in principle be caused by the anisotropy of the environment. However, then one would expect the NE component to be brighter due to a more efficient conversion of kinetic to radiative power \citep[e.g.][]{Orienti:2012fk,Wu:2013fk}. On the other hand, the difference in distances could possibly be due to light travel time effects that cause an apparent unequal ratio between the approaching and receding component, assuming intrinsically symmetric jet propagation. So, utilising the arm-length ratio of $R=1.4$ between the approaching and receding jet, we find that $\beta \cos \phi=(R-1)/(R+1)=0.17$, where $\beta$ is the velocity of the plasma blob in units of the speed of light, $c$, and $\phi$ is the angle between the approaching jet direction and the line of sight. The constraint of $\cos \phi \leq 1$ implies a lower limit to the velocity of $\beta \geq 0.17$ that is typical for hot spot advance speeds \citep{Liu:1992ve}, while $\beta \leq 1$ limits the viewing angle to $\phi \leq 80^{\circ}$. 

Considering the intrinsic diameters of the outer components and taking the distance between the central and outer components, we estimate the jet half opening angle to be $\sim11^{\circ}$ and a relativistic Mach number \citep{Falcke:1995yb} of the outer components as $M\simeq5$. Also, the lateral expansion velocity of the outer components would be $\sim0.2 c$ for a bulk velocity of $c$ along the jet axis. Doppler boosting of the central component may be negligible as a larger $\beta$ would result in a Doppler factor\footnote{$\delta=\left[\Gamma(1-\beta \cos \phi)\right]^{-1}$, where the Lorentz factor is $\Gamma = (1-\beta^{2})^{-0.5}$} below 1.18 for a fixed $\beta \cos \phi=0.17$, which would further increase its intrinsic brightness relative to the outer components.

Figure \ref{nebula} shows the associated large-scale radio bubble of Ho~II~X-1 together with the fine-scale radio structure demonstrating that the synchrotron jets fill the bubble. The main source of ionisation is the high X-ray luminosity, reaching $3\times10^{40}$~erg~s$^{-1}$ \citep{Grise:2010jf}. The lack of strong shock-ionisation \citep{Lehmann:2005bu,Egorov:2013ys} in the environment may indicate that the SW and NE components are not yet decelerated to their internal sound speed and are being decelerated possibly due to ram pressure \citep{Bureau:2002dq}. The lack of strong shocks could also indicate that the jet is young and has just started to drill into the ISM. Alternatively, the jet may not be continuous or persistent. We also note that there is marginal evidence for extended X-ray emission \citep{Lehmann:2005bu} that could be associated with the relativistic material. The central source was also observed to show hard and soft- X-ray spectra at a similar luminosity level \citep{Grise:2010jf}, which might be interpreted as transitions between super- and sub-Eddington regimes \citep[e.g.][]{Bachetti:2013uq}.

A naturally weighted image of Ho~II~X-1 (see the OM) shows diffuse emission between the components. So, the radio activity may not completely cease. The relative brightness ratio of the central component to the outer ones may indicate renewed radio activity. In this case, it could be that the outer components are from a former epoch of activity and the central component is from a more recent generation of ejecta. On the other hand, we may witness a semi-continuous jet, see in Sec 3.3.

The time elapsed between the appearance of this second generation of ejecta and a former phase of radio activity, the recurrence timescale ($t_{\rm{rec}}$), can be constrained from the morphology. To estimate the minimum light-travel time between the central and outer components, we take $\beta=1$ and $\sin 80^{\circ}$, which would imply $t_{\rm{rec}}=54-74$~yr, depending on which outer component is considered. On the other hand, taking into account that the outer components could be decelerated to $\sim0.17 c$ and using $\sin 10^{\circ}$ as the viewing angle  \footnote{If the source is not strongly Doppler boosted, this angle is a good lower limit to the viewing angle, even for a Lorentz factor as high as 15.}, we estimate a characteristic recurrence time of $t_{\rm{rec}}\simeq1790-2480$~yr, depending on the component. We note that these kinematic age estimates are not valid if the jet is continuous.

To compare timescales between relativistic sources, one may consider that for small viewing angles, $\phi<10^{\circ}$, light-travel times are expected to shorten by a factor of $2 \Gamma$. However, microblazars are expected to be rare \citep{Mirabel:1999bx} and the sources in the following comparison have larger viewing angles. Our derived recurrence time is at least an order of magnitude longer than the longest timescale of $\sim$4 years \citep{Corbel:2002qf} among microquasars where discrete ejecta were detected. In microquasars, multiple ejecta were detected on spatial scales of up to 5~pc \citep{Mirabel:1999bx,Fender:1999kx,Mirabel:1994bh}, which is also much less than the overall size of the triple morphology of 38.5~pc. On the other hand, flaring episodes typically last a fraction of the outburst duration. While the outburst durations are generally not well known in ULXs \citep{Burke:2013kl}, in microquasars they typically last from few months to $\sim3$ years. Even in GRS1915+105, which is known to be in an outburst since 1992, ejecta were detected up to $\sim3$ months \citep{Rushton:2010jy}.

A sign of restarted radio activity in supermassive BHs is a kiloparsec-size double-double or triple-double morphology \citep[e.g.][]{Schoenmakers:2000nx} with an age of about a billion years. On the other hand, VLBI monitoring campaigns have shown that ejecta on scales of tens of parsecs are indeed associated with young radio sources with intermittent activity; one of the youngest example has an age of 255~yr \citep{Wu:2013fk}. 

\subsection{Jet power and mass}

Detecting jets from Ho~II~X-1 allows us to constrain a time-averaged jet power associated with a photoionized environment of a ULX, where time-averaged X-ray outputs are well estimated from calorimetry \citep{Pakull:2002fk,Kaaret:2004ta,Berghea:2010vn}. To estimate the total jet power, we take the total radio luminosity of all the components, $(2.26\pm0.09)\times10^{34}$~erg~s$^{-1}$, as the central component is also optically thin. To obtain a conservative estimate, we use a scaling relationship that is based on minimum energy condition versus radio luminosity instead of e.g. cavity power versus radio luminosity. The latter would result in a somewhat higher jet power estimate and might be less suitable for Ho~II~X-1, see the OM for more details. We use a relationship that is obtained from a sample of FR~II sources of $Q_{\rm{j}}=(1.5\pm0.5)\times10^{44}\left(\frac{L_{151}}{10^{25} \rm{\,W\, Hz^{-1}\, sr^{-1}}}\right)^{0.67\pm0.05}$, where $L_{151}$ is the total radio luminosity at 151~MHz \citep{Godfrey:2013oq}. We estimate the radio power at 151~MHz using a radio spectral index of $\alpha=-0.8\pm0.2$ and find that $Q_{\rm{j}}=2.1^{+7.9}_{-1.8} \times10^{39}$~erg~s$^{-1}$. 

We recall that the isotropic X-ray luminosity needed to ionise the surrounding He {\sc ii} bubble is at least $4\times10^{39}$~erg s$^{-1}$ \citep{Kaaret:2004ta}. This X-ray luminosity corresponds to an average ionising rate over the past 3000~yr, which is comparable to the recombination time of the He {\sc ii}. Additionally, \citet{Berghea:2010vn} obtain a bolometric luminosity for Ho~II~X-1 of $L_{\rm{bol}}=1.34 \times 10^{40}$~erg~s$^{-1}$ by fitting the SED and modelling infrared lines, and argue that this estimate can be treated as time-averaged over at least 100~yr, based on the [O{\sc iv}] recombination time. The minimum isotropic X-ray luminosity truly exceeds the Eddington luminosity of a 25~M$_{\odot}$ BH as it is only a factor of ten less than the peak of the X-ray luminosity. So it is unlikely that the source is geometrically beamed. Moreover, the minimum total jet power already reaches the isotropic X-ray luminosity. Taking into account that a (time-averaged) jet power might only be as high as 10\% of the (time-averaged) bolometric luminosity \citep{Falcke:1995yb,Russell:2007ix}, this suggests that Ho~II~X-1 truly accretes, on average, at a high Eddington rate with a minimum BH mass of 25~M$_{\odot}$. Taking $L_{\rm{bol}}$ at face value or decreasing the jet efficiency would require an even higher BH mass.

\subsection{Active phase and environment}

The location and the size of the He {\sc ii} nebula and of the triple radio structure are well matched (see Fig. \ref{overlay}.), which may indicate the coupling of accretion and ejection. The surrounding radio bubble could then be sustained by Ho~II~X-1 in a manner similar to the Perseus cluster, where continuous bubble formation by the jets is thought to be the source that balances radiative cooling \citep{Fabian:2003cr}. In this regard, it is expected that both the overall radio spectral index of the bubble, $\alpha=-0.53\pm0.07$ \citep{Cseh:2012fk}, and of the inner structure is optically thin. Multiple episodes of relativistic ejecta could possibly result in the formation of a second (or third) generation of hot spots like in double-double (DDRGs) or triple-double radio galaxies \citep{Schoenmakers:2000nx} or in young compact symmetric objects (CSOs) \citep{Wu:2013fk}, and may diffuse away once decelerated below a Mach number of unity. However, it is yet to be explained what mechanism can drive an episodic behaviour of an accreting object. 

GRS1915+105 is thought to be wobbling across the ``jet line" and observed to show superluminal ejections associated with X-ray state transitions. These states and transitions could differ not just from the canonical ones -- e.g. hard state vs. plateau state  \citep{van-Oers:2010fk} -- but also from the ones Ho~II~X-1 has, and it is unclear how to reconcile such phenomenology with CSOs and DDRGs. 

On the other hand, the peculiar source SS433 also shows radio active and quiescent periods, but its jet consists of continuous ejecta \citep[e.g.][]{Paragi:1999xd}. So, if we are viewing a continuous jet of Ho~II~X-1, then SS433 is a better analogue than GRS1915+105. To alter the picture, we note that simulations by \citet{Goodall:2011kx} suggest the need of intermittency in the jet activity and a jet lifetime of $\sim$1600~yr for SS433 to perturb its SNR. Despite there are many similarities between all of these sources in radio, their time-dependent X-ray behaviour needs more investigation. In general, it is difficult to determine the presence of any recurrent activity as signatures of past activity may gradually get diffused away, precluding us from observing multiple phases of renewed activity. 

However, due to the long synchrotron cooling time, the radio bubble of Ho~II~X-1 possibly preserved any past radio activity. This may allow us to constrain the active lifetime of the source. The minimum total energy stored in the radio bubble is $E=2.6\times10^{49}$~erg, assuming no proton acceleration \citep{Miller:2005xh,Cseh:2012fk}. So, with a time-averaged total jet power of $Q_{\rm{j}}=2.1\times10^{39}$~erg~s$^{-1}$ this energy can be deposited into the environment over a timescale of $t=\frac{E}{Q_{\rm{j}}}=390$~yr. This suggests a relatively short active lifetime of the source.

Our results strengthen the view that physical properties of accretion and ejection are scale invariant over a possibly homogeneously populated BH mass range. Future studies may confirm a distinct formation channel of massive stellar-mass BHs, that are possibly caught for a short active time and evolve fast in environments akin to early cosmological conditions.

\section*{Acknowledgments}
The National Radio Astronomy Observatory is a facility of the National Science Foundation operated under cooperative agreement by Associated Universities, Inc. 
ZP and SF acknowledge support from the Hungarian Scientific Research Fund (OTKA K104539). SAF is the recipient of an Australian Research Council Postdoctoral Fellowship, funded by grant DP110102889.

\bibliographystyle{mn2e} 

{\small \bibliography{my}}

\begin{thebibliography}{}

\bibitem[\protect\citeauthoryear{{Bachetti}, {Rana}, {Walton}, {Barret},
  {Harrison}, {Boggs}, {Christensen} \& {Craig}}{{Bachetti}
  et~al.}{2013}]{Bachetti:2013uq}
{Bachetti} M. et al., 2013, ArXiv e-prints,
  2013arXiv1310.0745B

\bibitem[\protect\citeauthoryear{{Bauer} \& {Brandt}}{{Bauer} \&
  {Brandt}}{2004}]{Bauer:2004sh}
{Bauer} F.~E.,  {Brandt} W.~N.,  2004, ApJL, 601, L67

\bibitem[\protect\citeauthoryear{{Belczynski}, {Bulik}, {Fryer}, {Ruiter},
  {Valsecchi}, {Vink} \& {Hurley}}{{Belczynski}
  et~al.}{2010}]{Belczynski:2010kx}
{Belczynski} K.,  {Bulik} T.,  {Fryer} C.~L.,  {Ruiter} A.,  {Valsecchi} F.,
  {Vink} J.~S.,    {Hurley} J.~R.,  2010, ApJ, 714, 1217

\bibitem[\protect\citeauthoryear{{Berghea}, {Dudik}, {Weaver} \&
  {Kallman}}{{Berghea} et~al.}{2010}]{Berghea:2010vn}
{Berghea} C.~T.,  {Dudik} R.~P.,  {Weaver} K.~A.,    {Kallman} T.~R.,  2010,
  ApJ, 708, 364

\bibitem[\protect\citeauthoryear{{Bureau} \& {Carignan}}{{Bureau} \&
  {Carignan}}{2002}]{Bureau:2002dq}
{Bureau} M.,  {Carignan} C.,  2002, AJ, 123, 1316

\bibitem[\protect\citeauthoryear{{Burke}, {Kraft}, {Soria}, {Maccarone},
  {Raychaudhury}, {Sivakoff}, {Birkinshaw}, {Brassington}, {Forman},
  {Hardcastle}, {Jones}, {Murray} \& {Worrall}}{{Burke}
  et~al.}{2013}]{Burke:2013kl}
{Burke} M.~J. et al., 2013, ApJ, 775, 21

\bibitem[\protect\citeauthoryear{{Cavagnolo}, {McNamara}, {Nulsen}, {Carilli},
  {Jones} \& {B{\^\i}rzan}}{{Cavagnolo} et~al.}{2010}]{Cavagnolo:2010ly}
{Cavagnolo} K.~W.,  {McNamara} B.~R.,  {Nulsen} P.~E.~J.,  {Carilli} C.~L.,
  {Jones} C.,    {B{\^\i}rzan} L.,  2010, ApJ, 720, 1066

\bibitem[\protect\citeauthoryear{{Corbel}, {Fender}, {Tzioumis}, {Tomsick},
  {Orosz}, {Miller}, {Wijnands} \& {Kaaret}}{{Corbel}
  et~al.}{2002}]{Corbel:2002qf}
{Corbel} S.,  {Fender} R.~P.,  {Tzioumis} A.~K.,  {Tomsick} J.~A.,  {Orosz}
  J.~A.,  {Miller} J.~M.,  {Wijnands} R.,    {Kaaret} P.,  2002, Science, 298,
  196

\bibitem[\protect\citeauthoryear{{Crowther}, {Barnard}, {Carpano}, {Clark},
  {Dhillon} \& {Pollock}}{{Crowther} et~al.}{2010}]{Crowther:2010fk}
{Crowther} P.~A.,  {Barnard} R.,  {Carpano} S.,  {Clark} J.~S.,  {Dhillon}
  V.~S.,    {Pollock} A.~M.~T.,  2010, MNRAS, 403, L41

\bibitem[\protect\citeauthoryear{{Cseh}, {Corbel}, {Kaaret}, {Lang},
  {Gris{\'e}}, {Paragi}, {Tzioumis}, {Tudose} \& {Feng}}{{Cseh}
  et~al.}{2012}]{Cseh:2012fk}
{Cseh} D. et al., 2012, ApJ, 749, 17

\bibitem[\protect\citeauthoryear{{Egorov}, {Lozinskaya} \& {Moiseev}}{{Egorov}
  et~al.}{2013}]{Egorov:2013ys}
{Egorov} O.~V.,  {Lozinskaya} T.~A.,    {Moiseev} A.~V.,  2013, MNRAS, 429,
  1450

\bibitem[\protect\citeauthoryear{{Fabian}, {Sanders}, {Allen}, {Crawford},
  {Iwasawa}, {Johnstone}, {Schmidt} \& {Taylor}}{{Fabian}
  et~al.}{2003}]{Fabian:2003cr}
{Fabian} A.~C.,  {Sanders} J.~S.,  {Allen} S.~W.,  {Crawford} C.~S.,  {Iwasawa}
  K.,  {Johnstone} R.~M.,  {Schmidt} R.~W.,    {Taylor} G.~B.,  2003, MNRAS,
  344, L43

\bibitem[\protect\citeauthoryear{{Falcke} \& {Biermann}}{{Falcke} \&
  {Biermann}}{1995}]{Falcke:1995yb}
{Falcke} H.,  {Biermann} P.~L.,  1995, A\&A, 293, 665

\bibitem[\protect\citeauthoryear{{Falcke}, {K{\"o}rding} \& {Markoff}}{{Falcke}
  et~al.}{2004}]{Falcke:2004wm}
{Falcke} H.,  {K{\"o}rding} E.,    {Markoff} S.,  2004, A\&A, 414, 895

\bibitem[\protect\citeauthoryear{{Fender}, {Garrington}, {McKay}, {Muxlow},
  {Pooley}, {Spencer}, {Stirling} \& {Waltman}}{{Fender}
  et~al.}{1999}]{Fender:1999kx}
{Fender} R.~P.,  {Garrington} S.~T.,  {McKay} D.~J.,  {Muxlow} T.~W.~B.,
  {Pooley} G.~G.,  {Spencer} R.~E.,  {Stirling} A.~M.,    {Waltman} E.~B.,
  1999, MNRAS, 304, 865

\bibitem[\protect\citeauthoryear{{Fender}, {Homan} \& {Belloni}}{{Fender}
  et~al.}{2009}]{Fender:2009bh}
{Fender} R.~P.,  {Homan} J.,    {Belloni} T.~M.,  2009, MNRAS, 396, 1370

\bibitem[\protect\citeauthoryear{{Fryer}}{{Fryer}}{1999}]{Fryer:1999fk}
{Fryer} C.~L.,  1999, ApJ, 522, 413

\bibitem[\protect\citeauthoryear{{Gallo}, {Fender}, {Kaiser}, {Russell},
  {Morganti}, {Oosterloo} \& {Heinz}}{{Gallo} et~al.}{2005}]{Gallo:2005wa}
{Gallo} E.,  {Fender} R.,  {Kaiser} C.,  {Russell} D.,  {Morganti} R.,
  {Oosterloo} T.,    {Heinz} S.,  2005, Nature, 436, 819

\bibitem[\protect\citeauthoryear{{Goad}, {Roberts}, {Reeves} \&
  {Uttley}}{{Goad} et~al.}{2006}]{Goad:2006fk}
{Goad} M.~R.,  {Roberts} T.~P.,  {Reeves} J.~N.,    {Uttley} P.,  2006, MNRAS,
  365, 191

\bibitem[\protect\citeauthoryear{{Godfrey} \& {Shabala}}{{Godfrey} \&
  {Shabala}}{2013}]{Godfrey:2013oq}
{Godfrey} L.~E.~H.,  {Shabala} S.~S.,  2013, ApJ, 767, 12

\bibitem[\protect\citeauthoryear{{Goodall}, {Blundell} \& {Bell
  Burnell}}{{Goodall} et~al.}{2011}]{Goodall:2011kx}
{Goodall} P.~T.,  {Blundell} K.~M.,    {Bell Burnell} S.~J.,  2011, MNRAS, 414,
  2828

\bibitem[\protect\citeauthoryear{{Gris{\'e}}, {Kaaret}, {Feng}, {Kajava} \&
  {Farrell}}{{Gris{\'e}} et~al.}{2010}]{Grise:2010jf}
{Gris{\'e}} F.,  {Kaaret} P.,  {Feng} H.,  {Kajava} J.~J.~E.,    {Farrell}
  S.~A.,  2010, ApJL, 724, L148

\bibitem[\protect\citeauthoryear{{Justham} \& {Schawinski}}{{Justham} \&
  {Schawinski}}{2012}]{Justham:2012vn}
{Justham} S.,  {Schawinski} K.,  2012, MNRAS, 423, 1641

\bibitem[\protect\citeauthoryear{{Kaaret}, {Ward} \& {Zezas}}{{Kaaret}
  et~al.}{2004}]{Kaaret:2004ta}
{Kaaret} P.,  {Ward} M.~J.,    {Zezas} A.,  2004, MNRAS, 351, L83

\bibitem[\protect\citeauthoryear{{Karachentsev}, {Dolphin}, {Geisler},
  {Grebel}, {Guhathakurta}, {Hodge}, {Karachentseva}, {Sarajedini}, {Seitzer}
  \& {Sharina}}{{Karachentsev} et~al.}{2002}]{Karachentsev:2002fk}
{Karachentsev} I.~D. et al. 2002, A\&A, 383, 125

\bibitem[\protect\citeauthoryear{{K{\"o}rding}, {Jester} \&
  {Fender}}{{K{\"o}rding} et~al.}{2006}]{Kording:2006fk}
{K{\"o}rding} E.~G.,  {Jester} S.,    {Fender} R.,  2006, MNRAS, 372, 1366

\bibitem[\protect\citeauthoryear{{Lehmann}, {Becker}, {Fabrika}, {Roth},
  {Miyaji}, {Afanasiev}, {Sholukhova}, {S{\'a}nchez}, {Greiner}, {Hasinger},
  {Costantini}, {Surkov} \& {Burenkov}}{{Lehmann}
  et~al.}{2005}]{Lehmann:2005bu}
{Lehmann} I. et al., 2005, A\&A,
  431, 847

\bibitem[\protect\citeauthoryear{{Liu}, {Pooley} \& {Riley}}{{Liu}
  et~al.}{1992}]{Liu:1992ve}
{Liu} R.,  {Pooley} G.,    {Riley} J.~M.,  1992, MNRAS, 257, 545

\bibitem[\protect\citeauthoryear{{Mapelli}}{{Mapelli}}{2007}]{Mapelli:2007uq}
{Mapelli} M.,  2007, MNRAS, 376, 1317

\bibitem[\protect\citeauthoryear{{Mapelli} \& {Bressan}}{{Mapelli} \&
  {Bressan}}{2013}]{Mapelli:2013uq}
{Mapelli} M.,  {Bressan} A.,  2013, MNRAS, 430, 3120

\bibitem[\protect\citeauthoryear{{McHardy}, {Koerding}, {Knigge}, {Uttley} \&
  {Fender}}{{McHardy} et~al.}{2006}]{McHardy:2006zk}
{McHardy} I.~M.,  {Koerding} E.,  {Knigge} C.,  {Uttley} P.,    {Fender} R.~P.,
   2006, Nature, 444, 730

\bibitem[\protect\citeauthoryear{{Merloni}, {Heinz} \& {di Matteo}}{{Merloni}
  et~al.}{2003}]{Merloni:2003ri}
{Merloni} A.,  {Heinz} S.,    {di Matteo} T.,  2003, MNRAS, 345, 1057

\bibitem[\protect\citeauthoryear{{Mezcua}, {Roberts}, {Sutton} \&
  {Lobanov}}{{Mezcua} et~al.}{2013}]{Mezcua:2013fk}
{Mezcua} M.,  {Roberts} T.~P.,  {Sutton} A.~D.,    {Lobanov} A.~P.,  2013,
  ArXiv e-prints, 2013arXiv1309.5721M

\bibitem[\protect\citeauthoryear{{Miller}, {Mushotzky} \& {Neff}}{{Miller}
  et~al.}{2005}]{Miller:2005xh}
{Miller} N.~A.,  {Mushotzky} R.~F.,    {Neff} S.~G.,  2005, ApJL, 623, L109

\bibitem[\protect\citeauthoryear{{Mirabel}, {Dijkstra}, {Laurent}, {Loeb} \&
  {Pritchard}}{{Mirabel} et~al.}{2011}]{Mirabel:2011kx}
{Mirabel} I.~F.,  {Dijkstra} M.,  {Laurent} P.,  {Loeb} A.,    {Pritchard}
  J.~R.,  2011, A\&A, 528, A149

\bibitem[\protect\citeauthoryear{{Mirabel} \& {Rodr{\'{\i}}guez}}{{Mirabel} \&
  {Rodr{\'{\i}}guez}}{1994}]{Mirabel:1994bh}
{Mirabel} I.~F.,  {Rodr{\'{\i}}guez} L.~F.,  1994, Nature, 371, 46

\bibitem[\protect\citeauthoryear{{Mirabel} \& {Rodr{\'{\i}}guez}}{{Mirabel} \&
  {Rodr{\'{\i}}guez}}{1999}]{Mirabel:1999bx}
{Mirabel} I.~F.,  {Rodr{\'{\i}}guez} L.~F.,  1999, ARAA, 37, 409

\bibitem[\protect\citeauthoryear{{Orienti} \& {Dallacasa}}{{Orienti} \&
  {Dallacasa}}{2012}]{Orienti:2012fk}
{Orienti} M.,  {Dallacasa} D.,  2012, MNRAS, 424, 532

\bibitem[\protect\citeauthoryear{{Pakull} \& {Mirioni}}{{Pakull} \&
  {Mirioni}}{2002}]{Pakull:2002fk}
{Pakull} M.~W.,  {Mirioni} L.,  2002, ArXiv Astrophysics e-prints, 0202488

\bibitem[\protect\citeauthoryear{{Pakull}, {Soria} \& {Motch}}{{Pakull}
  et~al.}{2010}]{Pakull:2010kq}
{Pakull} M.~W.,  {Soria} R.,    {Motch} C.,  2010, Nature, 466, 209

\bibitem[\protect\citeauthoryear{{Paragi}, {Vermeulen}, {Fejes}, {Schilizzi},
  {Spencer} \& {Stirling}}{{Paragi} et~al.}{1999}]{Paragi:1999xd}
{Paragi} Z.,  {Vermeulen} R.~C.,  {Fejes} I.,  {Schilizzi} R.~T.,  {Spencer}
  R.~E.,    {Stirling} A.~M.,  1999, A\&A, 348, 910

\bibitem[\protect\citeauthoryear{{Prestwich}, {Kilgard}, {Crowther}, {Carpano},
  {Pollock}, {Zezas}, {Saar}, {Roberts} \& {Ward}}{{Prestwich}
  et~al.}{2007}]{Prestwich:2007kx}
{Prestwich} A.~H. et al., 2007, ApJL, 669, L21

\bibitem[\protect\citeauthoryear{{Rushton}, {Spencer}, {Fender} \&
  {Pooley}}{{Rushton} et~al.}{2010}]{Rushton:2010jy}
{Rushton} A.,  {Spencer} R.,  {Fender} R.,    {Pooley} G.,  2010, A\&A, 524,
  A29

\bibitem[\protect\citeauthoryear{{Russell}, {Fender}, {Gallo} \&
  {Kaiser}}{{Russell} et~al.}{2007}]{Russell:2007ix}
{Russell} D.~M.,  {Fender} R.~P.,  {Gallo} E.,    {Kaiser} C.~R.,  2007, MNRAS,
  376, 1341

\bibitem[\protect\citeauthoryear{{Schoenmakers}, {de Bruyn}, {R{\"o}ttgering},
  {van der Laan} \& {Kaiser}}{{Schoenmakers}
  et~al.}{2000}]{Schoenmakers:2000nx}
{Schoenmakers} A.~P.,  {de Bruyn} A.~G.,  {R{\"o}ttgering} H.~J.~A.,  {van der
  Laan} H.,    {Kaiser} C.~R.,  2000, MNRAS, 315, 371

\bibitem[\protect\citeauthoryear{{Silverman} \& {Filippenko}}{{Silverman} \&
  {Filippenko}}{2008}]{Silverman:2008uq}
{Silverman} J.~M.,  {Filippenko} A.~V.,  2008, ApJL, 678, L17

\bibitem[\protect\citeauthoryear{{van Oers}, {Markoff}, {Rahoui}, {Maitra},
  {Nowak}, {Wilms}, {Castro-Tirado}, {Rodriguez}, {Dhawan} \& {Harlaftis}}{{van
  Oers} et~al.}{2010}]{van-Oers:2010fk}
{van Oers} P. et al., 2010, MNRAS, 409, 763

\bibitem[\protect\citeauthoryear{{Volonteri}}{{Volonteri}}{2012}]{Volonteri:2012uq}
{Volonteri} M.,  2012, Science, 337, 544

\bibitem[\protect\citeauthoryear{{Wu}, {An}, {Baan}, {Hong}, {Stanghellini},
  {Frey}, {Xu}, {Liu} \& {Wang}}{{Wu} et~al.}{2013}]{Wu:2013fk}
{Wu} F. et al., 2013, A\&A, 550, A113

\bibitem[\protect\citeauthoryear{{Zampieri} \& {Roberts}}{{Zampieri} \&
  {Roberts}}{2009}]{Zampieri:2009if}
{Zampieri} L.,  {Roberts} T.~P.,  2009, MNRAS, 400, 677

\end{thebibliography}
\label{lastpage}

\clearpage
\onecolumn
\section*{Online Material}
Given that Ho~II~X-1 is associated with a radio and optical bubble, one may use a relationship between cavity power and radio power derived from a sample of low-power AGNs, typically FR~Is, to estimate the jet power. We use $\log Q_{\rm{j}}=0.75 \log P_{1.4}+1.91$ that has a scatter of 0.78~dex and where $Q_{\rm{j}}$ is in units of $10^{42}$~erg~s$^{-1}$ and the radio power at 1.4~GHz is in units of  $10^{40}$~erg~s$^{-1}$ (Cavagnolo 2010). After substituting the total radio power at 1.4~GHz, which we obtained using a radio spectral index of $\alpha=-0.8$, we find that $Q_{\rm{j}}=1.36\times10^{40}$~erg~s$^{-1}$ with an uncertainty of a factor of six. This is higher than the estimate in Sec. 3. but consistent within the errors. On the other hand, from the apparent morphology of Ho~II~X-1 it follows that it is more similar to FR~IIs. An analogy with FR~IIs, instead of FR~Is, is expected on the ground of a higher accretion state (K\"{o}rding 2006).  

\begin{table*}
\caption{Summary of results \label{fit}}
\begin{tabular}{cccccc}
\hline\hline
Observation&Component&Peak intensity&Flux density&Component size$^a$&Central\\
date&&[$\mu$Jy/b]&[$\mu$Jy]&[mas $\times$ mas]$^b$&frequency [GHz]\\
\hline
16 Nov 12&Central&$145\pm3$&$172\pm6$&--&5.24\\
17 Nov 12&Central&$160\pm3$&$175\pm6$&--&5.24\\
\hline
Combined&Central&$151\pm2$&$174\pm4$&--&5.24\\
&SW&$25\pm2$&$85\pm9$&$(570\pm60) \times (520 \pm 50)$&5.24\\
&NE&$23\pm2$&$55\pm7$&$(520\pm60) \times (290\pm40)$&5.24\\
\hline
Combined&Central&$162\pm4$&$197\pm8$&--&4.75\\
&SW&$32\pm4$&$97\pm15$&$(640\pm90) \times (500 \pm 70)$&4.75\\
&NE&$31\pm4$&$62\pm11$&$(530\pm80) \times (250\pm60)$&4.75\\
\hline
Combined&Central&$140\pm4$&$171\pm8$&--&5.75\\
&SW&$24\pm4$&$82\pm16$&$(610\pm110) \times (410 \pm 80)$&5.75\\
&NE&$24\pm4$&$52\pm11$&$(420\pm90) \times (290\pm60)$&5.75\\
\hline
\end{tabular}
\medskip \\$^a$Intrinsic size, deconvolved from beam. $^b$ mas denotes milli-arcsecond. The penultimate block shows the results of imaging the first four spectral windows. 
The last block shows the results of imaging the third four of the spectral windows.
\end{table*}

\begin{figure}
\begin{center}
\resizebox{4in}{!}{
\includegraphics{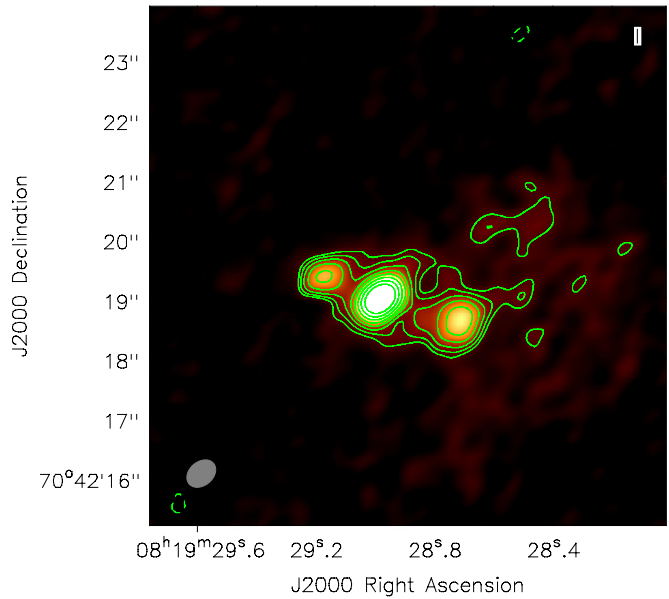}}
\caption{The naturally weighted C-band JVLA A-array image of Ho II X-1. The lowest contours represent radio emission at $\pm(\sqrt{2})^{3}$ times the rms noise level of 2.5~$\mu$Jy beam$^{-1}$. Contours are increased by $(\sqrt{2})^{n}$$\sigma$, where $n\in[4;10]$, and the peak brightness is 162~$\mu$Jy~beam$^{-1}$. The Gaussian restoring beam size is $0.53" \times 0.40"$ at a position angle of --50$^\circ$. (The background colour image, with smoothed pixels, represents radio emission on an arbitrary scale of $0-50$~$\mu$Jy to match the contours.)}
\label{natural}
\end{center}
\end{figure}

\end{document}